\documentclass[11pt]{article}

\usepackage[margin=.75in]{geometry}
\usepackage{datetime}
\usepackage{color}
\usepackage{amsmath, amsthm, amssymb,fancyhdr,mathrsfs,  enumerate}
\usepackage{latexsym}
\usepackage[round, authoryear, comma, sort&compress]{natbib}
\usepackage{hyperref}
\usepackage[utf8]{inputenc}
\usepackage{graphicx,float}
\usepackage[export]{adjustbox}

\newtheorem{lemma}{Lemma}

\newcommand{\bpsi}{{\bf\Psi}}
\newcommand{\bb}{{\bf\beta}}

\newcommand{\ba}{{\bf \alpha}}

\newcommand{\E}{\mathbb{E}}
\newcommand{\I}{\mathbb{I}}
\newcommand{\R}{\mathbb{R}}

\newcommand{\widesim}[2][1.5]{
  \mathrel{\overset{#2}{\scalebox{#1}[1]{$\sim$}}}
}

\begin{document}

\begin{center}
      \Large{\sc{Liu-type Shrinkage Estimators for Mixture of Poisson Regressions with Experts: A Heart Disease Study}}
\end{center}
\begin{center}
 \noindent{{\sc Elsayed Ghanem$^{\dagger,\ddagger}$}, 
            {\sc Moein Yoosefi}$^{\dagger}$} and
 {\sc Armin Hatefi$^{\dagger,}$\footnote{Corresponding author:\\
 Email: ahatefi@mun.ca and Tel: +1 (709) 864-8416}}
 
\vspace{0.5cm}
\noindent{\footnotesize{\it $^{\dagger}$Department of Mathematics and Statistics, Memorial University of Newfoundland, St. John's, NL, Canada.}} \\
\noindent{\footnotesize{\it $^{\ddagger}$Faculty of Science, Alexandria University, Arab Republic of Egypt.}}
\end{center}


\begin{center} {\small \bf Abstract}:
\end{center}
{\small
 Count data play a critical role in medical research, such as heart disease.
 The Poisson regression model is a common technique for evaluating the impact of a set of covariates on the count responses. The mixture of Poisson regression models with experts is a practical tool to exploit the covariates, not only to handle the heterogeneity in the Poisson regressions but also to learn the mixing structure of the population. Multicollinearity is one of the most common challenges with regression models, leading to ill-conditioned design matrices of Poisson regression components and expert classes. The maximum likelihood method produces unreliable and misleading estimates for the effects of the covariates in multicollinearity. In this research, we develop Ridge and Liu-type methods as two shrinkage approaches to cope with the ill-conditioned design matrices of the mixture of Poisson regression models with experts. Through various numerical studies, we demonstrate that the shrinkage methods offer more reliable estimates for the coefficients of the mixture model in multicollinearity while maintaining the classification performance of the ML method. The shrinkage methods are finally applied to a heart study to analyze the heart disease rate stages. 
}



\section{Introduction} \label{sec:intro}

 Cardiovascular diseases (CVDs) cause severe health issues by affecting the heart and blood vessels. When the blood supply to the brain is interrupted, the CVDs lead to a stroke and transient ischemic attacks. If the blood supply to the heart is impeded, angina and heart attacks occur, and CVDs result in coronary heart disease. Peripheral arterial disorders, in which fatty deposits in arteries limit blood flow to the arms and legs, are greatly increased by CVDs. Heart failure, heart valve disorders, and vascular dementia are just a few of the many cardiac diseases that CVDs can cause \citep{tsao2023heart,writing2012heart, cohn1997report, go2014heart}.

CVDs are among the top causes of mortality, disability, and immobility in the US, Canada, and the EU. For example, 1 in 3 deaths in the US occurs as a result of CVDs each year. Also, 22\% of all premature deaths in the UK are caused by CVDs. By 2030, it is anticipated that 40.5\% of the US population will experience some form of CVDs, with coronary heart disease growing by 18\% annually and stroke growing by 50\% \cite{zeiger2012evaluation, andrade20202020, leal2006economic,writing2012heart,heidenreich2011forecasting}.   
The economic burden of the CVDs is also undeniable on healthcare systems. While the overall cost of all cancers and benign neoplasms was \$ 201.5 billion in 2008, the entire cost of CVDs and stroke accounted for \$314.5 billion in 2010 \cite{go2014heart}.
Therefore, it is necessary to monitor and build prognosis methods to determine the prevalence of heart diseases in the communities. This enables us to better inform the decision-makers of the heterogeneity in the disease population and how the risk factors vary in heterogeneous sub-populations.

There are various risk factors associated with heart disease. Due to physiological differences based on sex and age, male and female CVD patients typically behave differently. While the prevalence of heart disease typically increases in women after age 60, it strikes men earlier in life. Other comorbidities and risk factors that increase the risk of heart disease in addition to age and sex include, for example, hypertension, high cholesterol, physical activity, obesity, and irregular heart rhythm \cite{go2014heart, berry2012lifetime, stampfer2000primary}. Finding the risk factors and understanding how they impact heart-related functioning problems may provide comprehensive perspectives on managing and prognosis heart disorders.

Linear and generalized linear models are common statistical methods to study the impact of the risk factors on heart disease. While various methods have been developed in the literature to find the effect of covariates on the heart disease rate stages, Poisson regression models remain a flexible method to handle the heart disease data because of the essential structure of the counting responses \citep{prosser2007predictors,pranata2020impact, mufudza2016poisson}. When covariates are linearly dependent, one of the key problems with Poisson regression models is multicollinearity. The maximum likelihood (ML) estimates for the coefficients of regression produce unreliable and misleading estimates with significantly inflated variances in the presence of multicollinearity.

The ridge estimation method is a common and feasible approach to address the multicollinearity problem in Poisson regressions \citep{qasim2020new}. 
Despite this, when colinearity becomes high, the ridge approach might not be able to sufficiently encounter the ill-conditioned design matrix. Smaller values are insufficient to deal with the ill-conditioned design matrix, while large ridge parameter values appear to introduce substantial biases in the estimates. 
\cite{liu2003using} proposed Liu-type (LT) shrinkage method employing two tuning parameters to handle the ill-conditions design matrix. \cite{qasim2020new} applied the LT shrinkage estimation method to the Poisson regressions.  
\cite{arashi2014improved} applied the LT method to improve the preliminary tests and stein-rule for elliptical linear regressions. \cite{inan2013liu} proposed LT estimates for logistic regressions.
\cite{roozbeh2013feasible} applied the LT estimation method to partially linear regression models.  
\cite{pearce2021multiple} proposed rank-based LT estimates for logistic and stochastic regression models.

The probabilistic finite mixture models combine multiple distributions to more effectively model heterogeneous populations \citep{fruhwirth2019handbook}. 
The finite mixture of regressions is a practical method to incorporate a set of covariates in learning the heterogeneity of the regression components in an unsupervised approach. Maximum likelihood (ML) is a standard method to estimate the parameters of the mixture models.  
The expectation-maximization (EM) algorithm method contributes as the practical strategy to find the ML estimates of the parameters of the mixture of regression models.
\cite{celeux1985sem} developed the stochastic EM algorithm by adding a stochastic teacher to the EM algorithm in estimating the parameters of mixture models. \cite{faria2010fitting} applied the SEM algorithm to the mixture of linear regression models.  
\cite{shalabh2008finite} extended the mixture of the generalized linear models. \cite{mufudza2016poisson} proposed a mixture of Poisson regression models to analyze the heterogeneity in heart disease rates. See \cite{mclachlan2019finite,fruhwirth2019handbook} and references therein for more information about the theory and applications of mixture models.

According to the correlation between covariates in regressions, the ML method leads to unreliable estimates for the coefficients of the mixture of regression models in the presence of multicollinearity. \cite{ghanem2022unsupervised,ghanem2022liu} recently developed shrinkage methods to deal with multicollinearity in the mixture of logistic and regression models. In this manuscript, we focus on the mixture of Poisson regression models with multinomial experts where the collinearity problem impacts the estimators for the coefficients of both Poisson regression components and experts.   To overcome this challenge,  we develop Ridge and Liu-type shrinkage estimation methods for the coefficients of the mixture of Poisson regressions with experts in multicollinearity. 
We compare the estimation and classification performance of the shrinkage methods with their ML counterparts in estimating the coefficients of regression components and expert classes through various numerical studies.   We show that the developed shrinkage methods produce more reliable results in estimating the coefficients of the mixture models. Finally, we apply the proposed methods to analyze the heart disease study to assess the disease rate stages.

This manuscript is organized as follows. Section \ref{sec:stat}
 develops the shrinkage methods in estimating the parameters of the mixture of Poisson regressions with experts. 
 Section \ref{sec:sim} evaluates the estimation and classification of the developed methods in various settings. 
 Section \ref{sec:real} applied the methods to analyze a heart disease example. Finally, we present the summary and concluding remarks in Section \ref{sec:con}.

\section{Statistical Methods} \label{sec:stat}
Let ${\bf x}_i^\top = (x_{i,1},\ldots,x_{i,p})$ denote $p$ explanatory variables for $i$-th individual from a random sample of size $n$ where
${\bf X}^\top = ({\bf x}_1^\top,\ldots,{\bf x}_n^\top)^\top$ represents the design matrix with $\text{Rank}({\bf X})=p<n$.
Let ${\bf y} =(y_1,\ldots,y_n)$ denote the count responses of $n$ individuals in the sample. Thus, Poisson regression model $y_i \sim \text{Poi}(\mu({\bf x}_i))$ where $\mu({\bf x}_i) = \exp({\bf x}_i^\top \beta)$ is a common model to study the effect of explanatory variables on the response variable.
In this manuscript, we focus on the finite mixture of the Poisson regression models with experts (FMPRE).
The FMPRE as a generalization of the Poisson regression model is given by
\begin{align}\label{fmpr}
    H(y_i|{\bf x}_i, {\bf \omega}_i,{\bf \Psi}) = \sum_{j=1}^{J} 
    \pi_j({\bf \omega}_i,{\bf \alpha}) 
    \text{Poi}(y_i| \mu_j({\bf x}_i)),
\end{align}
 where $\text{Poi}(y_i| \mu_j({\bf x}_i))$ denotes the Poisson distribution and $\log(\mu_j({\bf x}_i)) = {\bf x}_i^\top \beta_j$, ${\bf \Psi}=({\bf \alpha},{\bf \beta})$ where ${\bf \alpha}=(\alpha_1,\ldots, \alpha_J)$ and ${\bf \beta}=(\beta_1,\ldots,\beta_J)$ denotes the vector of all unknown parameters of the mixture model, $y_i$, ${\bf x}_i$ and ${\bf \omega}_i$ represent the values of response, explanatory and concomitant variables for the $i$-th individual, respectively. Also, let
 ${\bf \omega}^\top=({\bf \omega}_1^\top,\ldots,{\bf \omega}_n^\top)^\top$ represent 
 the design matrix in the expert classes with $\text{Rank}({\bf \omega})=q<n$.
 Throughout this manuscript, we assume that the number of the components of the mixture model $J$ is fixed and priori known; however,  the component membership of the observations is unknown and, in an unsupervised approach, should be estimated from the model.
 In the FMPRE model \eqref{fmpr}, the effect of concomitant variables of the mixing of the components is typically assessed through a multinomial logit model as
 \begin{align}\label{multi_logit}
     \pi_j({\bf \omega}_i,{\bf \alpha}) =
     \frac{\exp({\bf \omega}_i^\top {\alpha}_j)}{\sum_{u=1}^{J} \exp({\bf \omega}_i^\top {\alpha}_{u})},
 \end{align}
where ${\bf \alpha}=(\alpha_1^\top,\ldots,\alpha_J^\top)$ with $\alpha_1^\top \equiv {\bf 0}$ as the reference group. 
From the FMPRE model \ref{fmpr}, the log-likelihood function of ${\bf \Psi}$ is given by
\begin{align}\label{ll_inc}
\ell({\bf \Psi}|{\bf y}) = \sum_{i=1}^{n} \sum_{j=1}^{J}  
\pi_j({\bf \omega}_i,{\bf \alpha}) \exp \left\{ - \mu_j({\bf x}_i) + y_i \log (\mu_j({\bf x}_i)) - \log( y_i!)\right\}
\end{align}
where the parameters $\mu_j({\bf x}_i)$ from the responses are incorporated into the model by a separate vector of coefficients ${\bf\beta}_j$ within each component of the FMPRE.
The log-likelihood function \eqref{ll_inc} is not tractable to obtain the maximum likelihood estimate (MLE) of ${\bf \Psi}$.
To do so, it is common that we view input data $({\bf X},{\bf y})$ as incomplete data. The incomplete data is then augmented with the latent variables to obtain the MLE by the expectation-maximization (EM) algorithm of \cite{dempster1977maximum}.   

For each ${\bf x}_i, i=1,\ldots,n$, let ${\bf Z}_i=(Z_{i1},\ldots,Z_{iJ})$ denote the latent variable indicating the component memberhip of $i$-th individual in the mixture such that 
  \begin{align*}
 Z_{ij} = \left\{
 \begin{array}{lc}
 1 & \text{if the $i$-th individual comes from the $j$-th component,}\\
 0 & o.w.,
 \end{array} \right.
 \end{align*}  
where ${\bf Z}_i \widesim{iid} \text{Multi}\left(1, \pi_1({\bf \omega}_i,{\bf \alpha}),\ldots, \pi_J({\bf \omega}_i,{\bf \alpha})\right)$. 
From the joint distribution of $({\bf Z}_i,y_i)$, 
 it is easy to show   ${\bf Z}_i|y_i \widesim{iid} \text{Multi}(1, \tau_{i1}(\bpsi),\ldots,\tau_{iJ}(\bpsi))$ where
 \begin{align}\label{tau_ml}
 \tau_{ij}(\bpsi) = \frac{\pi_j({\bf \omega}_i,{\bf \alpha}) e^{-\mu_j({\bf x}_i)} [\mu_j({\bf x}_i)]^{y_i}}{\sum_{u=1}^{J} \pi_u({\bf \omega}_i,{\bf \alpha}) e^{-\mu_u({\bf x}_i)} [\mu_u({\bf x}_i)]^{y_i}}.
  \end{align} 
  Let ${\mathcal D}=({\bf X},{\bf y},{\bf Z})$ denote the complete data. Thus,   the complete log-likelihood function of $\bpsi$ is given by 
 \begin{align}\label{ll_comp_ml}
 \ell_c(\bpsi) = \sum_{i=1}^{n} \sum_{j=1}^{J} z_{ij} \log\left\{\pi_j({\bf \omega}_i,{\bf \alpha}) \right\}
 + \sum_{i=1}^{n} \sum_{j=1}^{J} z_{ij} 
 \log\left\{ y_i \log(\mu_j({\bf x}_i)) - \mu_j({\bf x}_i) - \log(y_i!)\right\}.
\end{align}

One can use the EM algorithm \citep{dempster1977maximum} to obtain the MLE of ${\bpsi}$. The EM algorithm decomposes the estimation problem into the iterative expectation (E) and maximization (M) steps. 
\cite{celeux1985sem}  proposed the stochastic EM (SEM) algorithm accommodating a stochastic classification (S) step between E- and M- steps in each iteration. 
\cite{faria2010fitting} applied the SEM algorithm to the mixture of regression models to obtain the MLE of coefficients of the component regressions. In this section, we propose an SEM algorithm to estimate not only the coefficient of the Poisson regressions but also the coefficients of the expert classes. 

Let $\bpsi^{(0)}$ and $\bpsi^{(t)}$ denote the initial value and the update of $\bpsi$ from the $t$-th iteration, respectively.  
In the E-step of the $(t+1)$-th iteration,  we compute the expected value of  \eqref{ll_comp_ml} by
\begin{align*}
     Q({\bpsi},{\bpsi}^{(t)}) = \E\left(\ell_c(\bpsi)|{\mathcal D}, {\bpsi}\right)|_{{\bpsi}={\bpsi}^{(t)}} \propto 
    Q_1({\bf \alpha}, {\bpsi}^{(t)}) + Q_2({\bf \beta},{\bpsi}^{(t)})
\end{align*}
where 
\begin{align}\label{Q1}
   Q_1({\bf \alpha}, {\bpsi}^{(t)}) & \propto  
   \sum_{i=1}^{n} \sum_{j=1}^{J} \tau_{ij}(\bpsi^{(t)}) \log\left\{\pi_j({\bf \omega}_i,{\bf \alpha}) \right\},\\
\label{Q2}
   Q_2({\bf \beta},{\bpsi}^{(t)}) & \propto 
   \sum_{i=1}^{n} \sum_{j=1}^{J} \tau_{ij}(\bpsi^{(t)}) 
   \log\left\{ y_i \log(\mu_j({\bf x}_i)) - \mu_j({\bf x}_i) \right\},
\end{align}
with $\tau_{ij}(\bpsi^{(t)})$ from \eqref{tau_ml}. In the S-step, we develop partition ${\bf P}^{(t+1)}= \left\{ P_1^{(t+1)},\ldots, P_J^{(t+1)}\right\}$ on the sample space and partition subjects using a stochastic assignment based on the posterior membership probabilities. To do so, for each subject,  we first generate $ \Tilde{{\bf Z}}^{(t+1)}_i \widesim{iid} \text{Multi}\left(1,\tau_{i1}(\bpsi^{(t)}),\ldots, \tau_{iJ}(\bpsi^{(t)}) \right)$  and then partition the subject as
$
({\bf x}_i,y_i) \in P_j^{(t+1)}$ when  $\Tilde{{ Z}}^{(t+1)}_{ij} =1$ for $j=1,\ldots,J$ and  $i=1,\ldots,n$.
Let $n_j$ denote the number of individuals assigned into partition $P_j^{(t+1)}, j=1,\ldots,n$ where $\sum_{j=1}^{J} n_j=n$.
Note that if a partition becomes empty in an iteration, then the EM algorithm is stopped. 

In the M-step, we plan to maximize the conditional expectation of the complete log-likelihood to obtain ${\bpsi}^{(t+1)}$. 
From \eqref{Q1} and \eqref{Q2}, one can update ${\bf \beta}^{(t+1)}$ and ${\bf \alpha}^{(t+1)}$, by maximizing separately $Q_1({\bf \alpha}, {\bpsi}^{(t)})$ and $Q_2({\bf \beta},{\bpsi}^{(t)})$.
To obtain ${\bf \beta}^{(t+1)}$, we can reformulate the maximization of \eqref{Q2} to an iterative re-weighted least squares (IRWLS) estimation problem. It is easy to show that  ${\bf \beta}^{(t+1)}_j$, as a solution to the IRWLS using \eqref{Q2}, is updated by 
\begin{align}
    {\widehat{\bf \beta}}_j^{(t+1)} = 
    \left( {\bf X}_j^\top {\bf W}_j {\bf X}_j  \right)^{-1}
    {\bf X}_j^\top {\bf W}_j {\bf z}^*_j,
\end{align}
where ${\bf X}_j$ and ${\bf y}_j$ are the design matrix and response vector of the individuals allocated to partition $P_j^{(t+1)}$ with  ${\bf z}^*_j = \left\{ {\bf X}_j {\widehat{\bf \beta}}_j^{(t)} +  \frac{1}{\mu_j({\bf x}_i)} 
[{\bf y} - \mu_j({\bf x}_i)]\right\}$ and 
${\bf W}_j$  is a diagonal weight matrix with $i$-th diagonal element $\mu_j({\bf x}_i)$ for $i=1,\ldots,n_j; j=1,\ldots,J$.

To update the ${\widehat{\bf \alpha}}^{(t+1)}$, we can view \eqref{Q1} as a log-likelihood function of ${\bf \alpha}$ from the multinational logit regression when $\tau_{ij}(\bpsi^{(t)})$ is replaced by $\Tilde{{ Z}}^{(t+1)}_{ij}$ from the S-step; hence $\Tilde{{ \bf Z}}^{(t+1)}_{i}, i=1,\ldots, n$ play the role of the multinomial responses and experts represents the logit regression corresponding to each component of the mixture population.  It is well known that there is no closed form for the  solution to 
\eqref{Q1}, thus we can translate the maximization step to a coordinate descent algorithm where we implement the IRWLS sequentially to estimate the coefficients of the $j$-th expert $j=1,\ldots, J$ in each iteration of coordinate descent. 
Suppose ${\bf W}_j({\bf \alpha}^{(t)})$ denotes the diagonal weight matrix of size $n$ in the $j$-th expert. In the $(t+1)$-th iteration of the SEM algorithm, let ${\bf W}_j({\bf \alpha}^{(t)})$ and ${\bf U}_j({\bf \alpha}^{(t)})$ 
\begin{align} \label{w_alpha}
   {\bf W}_j({\bf \alpha}^{(t)}) = 
   - {\text diag} \left[ \pi_j({\bf \omega}_1,{\bf \alpha}^{(t)}) (1- \pi_j({\bf \omega}_1,{\bf \alpha}^{(t)})),\ldots, \pi_j({\bf \omega}_{n},{\bf \alpha}^{(t)}) (1- \pi_j({\bf \omega}_{n},{\bf \alpha}^{(t)}))
   \right]
\end{align}
\begin{align}\label{u_alpha}
    {\bf U}_j({\bf \alpha}^{(t)}) =
\left[\I(\Tilde{{Z}}^{(t+1)}_{1j} =1) - 
\pi_j({\bf \omega}_1,{\bf \alpha}^{(t)}),\ldots, 
\I(\Tilde{{Z}}^{(t+1)}_{n,j} =1) - 
\pi_j({\bf \omega}_{n},{\bf \alpha}^{(t)})
\right]^\top
\end{align}
Accordingly, one can easily show that 
\begin{align}
   {\widehat{\bf \alpha}_j^{(t+1)}} = \arg \min_{{\alpha}_j} 
   || {\bf v}_j({\bf \alpha}^{(t)}) - {\bf \omega}^* {\bf \alpha}_j||_2^2
\end{align}
where ${\bf \omega}^* = {\bf W}_j^{1/2}({\bf \alpha}^{(t)}) {\bf \omega}$ and 
${\bf v}_j({\bf \alpha}^{(t)}) = {\bf W}_j^{1/2}({\bf \alpha}^{(t)}) \left( 
{\bf \omega} {\bf \alpha}_j^{(t)} + {\bf W}_j^{-1}({\bf \alpha}^{(t)}) {\bf U}_j({\bf \alpha}^{(t)}) \right)$. 
From \citep{celeux1985sem,faria2010fitting}, although the point-wise convergence in the SEM algorithm is not guaranteed, the SEM results in a Markov chain fluctuating around the MLE at the stationary state of the chain. Hence, we alternate the E-, S- and M-steps until either the chain meets the stopping rule $|\ell(\bpsi^{(t+1)}|{\bf y})-\ell(\bpsi^{(t)}|{\bf y})|< \epsilon$  or reaches a pre-specified maximum number of iterations fixed for all the estimation methods for a fair comparison.

\subsection{Ridge Estimation for FMPRE}\label{sub:ridge}  
The ML estimate discussed earlier is considered the most common approach in estimating the parameters of the FMPRE \eqref{fmpr}. 
Despite this popularity, the ML estimates are not robust against the multicollinearity in the design matrices.  In multicollinearity, the ML method appears unreliable and even misleading in estimating the coefficients of the component regressions and the expert classes. 
The ridge estimation method, proposed by \cite{hoerl1970ridge}, is an established remedy to handle the multicollinearity issues involved in the design matrix of the regression models.  Here, we develop the ridge estimates for the parameters of the FMPRE model \eqref{fmpr}.
One can derive the ridge estimate of ${\bpsi}$, as 
a solution to the log-likelihood function subject to the ridge penalty given by
\begin{align}\label{ll_inc_ridge}
    \ell^R({\bpsi}|{\bf y}) = \ell({\bpsi}|{\bf y}) - \lambda \bpsi^\top \bpsi /2,
\end{align}
where  $\bpsi=({\bf\alpha},{\bf\beta})$ is the vector of all unknown parameters and $\ell({\bpsi}|{\bf y})$ comes form \eqref{ll_inc}.

The incomplete ridge likelihood \eqref{ll_inc_ridge} is not tractable with respect to estimating the FMPRE  parameters. Similar to the ML approach,  we employ the missing data mechanism and introduce latent variables ${\bf Z}_i=(Z_{i1},\ldots, Z_{iJ})$ to accommodate the component membership of the incomplete observations $({\bf x}_i,y_i)$  for $i=1,\ldots,n$.
Let ${\mathcal D}=({\bf X},{\bf y},{\bf Z})$ denote the complete data. Thus, we propose the SEM algorithm to estimate the coefficients of the mixture model.
Let $\bpsi^{(0)}$ and $\bpsi^{(t)}$ represent, respectively, the initial point and the update from the $t$-th iteration of the SEM algorithm under the ridge method.  
Like the E-step of the ML estimation method, we compute the conditional expectation of the latent variables given the incomplete data and decompose the conditional expectation of the log-likelihood function to \eqref{Q1} and \eqref{Q2}. 

In the S-step of the $(t+1)$-th iteration, using the conditional expectation of the latent variables \eqref{tau_ml}, we again apply the stochastic step and generate  
$\Tilde{{\bf Z}}^{(t+1)}_i \widesim{iid} \text{Multi}\left(1,\tau_{i1}(\bpsi^{(t)}),\ldots, \tau_{iJ}(\bpsi^{(t)}) \right)$ for each observation $({\bf x}_i,y_i)$  for $i=1,\ldots,n$.
We then partition the data into ${\bf P}^{(t+1)}= \left\{ P_1^{(t+1)},\ldots, P_J^{(t+1)}\right\}$ and assign  $
({\bf x}_i,y_i) \in P_j^{(t+1)}$ when  $\Tilde{{ Z}}^{(t+1)}_{ij} =1$ for $j=1,\ldots,J$ and  $i=1,\ldots,n$. 
In the M-step, the regression coefficients ${\bf\beta}_j, j=1,\ldots, J$ are then obtained from Lemma \ref{lem:beta_ridge} through the IRWLS using the ridge penalty. 
\begin{lemma}\label{lem:beta_ridge}
    Under the assumption of FMPRE \eqref{fmpr}, the Poisson regression coefficients ${\bf\beta}_j, j=1,\ldots,J$ is updated, as a solution to the conditional expectation of log-likelihood $Q_2({\bf \beta},{\bpsi}^{(t)})$ under ridge penalty, by
    \[
    {\widehat{\bf\beta}}_j^{(t+1)} = 
    \left( {\bf X}_j^\top {\bf W}_j {\bf X}_j + \lambda_j \I \right)^{-1}
    {\bf X}_j^\top {\bf W}_j {\bf z}_j^*,
    \]
    where ${\bf W}_j$ is diagonal weight matrix of size $n_j$ with $i$-th diagonal entry $\mu_j({\bf x}_i)$ and 
    \[{\bf z}^*_j = \left\{ {\bf X}_j {\widehat{\bf \beta}}_j^{(t)} +  \frac{1}{\mu_j({\bf x}_i)} 
     [{\bf y} - \mu_j({\bf x}_i)]\right\}.
     \]
\end{lemma}
To update the coefficients of the experts, ${\bf\alpha}_j, j=1,\ldots,J$, 
one can focus on the expected log-likelihood \eqref{Q1} and replace 
${\tau}_{ij}(\bpsi)$
with their stochastic realizations $\Tilde{{Z}}^{(t+1)}_{ij}, \forall i,j$ from the S-step. The resulting log-likelihood can be viewed as the log-likelihood from the multinomial logit regression. 
Hence, ${\bf\alpha}_j$ under ridge penalty can be updated through the ridge penalized log-likelihood function of the multinomial logit regression.
There is no closed-form solution to this penalized log-likelihood function. We apply the coordinate decent method to numerically maximize the penalized log-likelihood function to obtain ${\widehat{\bf\alpha}}^{(t+1)}_j, j=1,\ldots, J$.
In each iteration of the coordinate decent, we apply Lemma \ref{lem:alpha_ridge} and iteratively update ${\widehat{\bf\alpha}}^{(t+1)}_j$ via IRWLS method using ridge penalty for $ j=1,\ldots, J$.
\begin{lemma}\label{lem:alpha_ridge}
    Under the assumption of FMPRE \eqref{fmpr}, the coefficients of the $j$-th expert ${\bf\alpha}_j$ under ridge penalty in the $(t+1)$-th iteration of the SEM is updated by
    \[
    {\widehat{\bf\alpha}}_j^{(t+1)} = 
    \left({\bf\omega}^\top {\bf W}_j({{\bf\alpha}}^{(t)}) {\bf\omega} + \lambda^*_j \I\right)^{-1} 
    {\bf\omega}^\top {\bf W}_j({{\bf\alpha}}^{(t)}) 
    {\bf v}_j({{\bf\alpha}}^{(t)}),
    \]
where  ${\bf W}_j({{\bf\alpha}}^{(t)})$ is computed by \eqref{w_alpha} and
${\bf v}_j({{\bf\alpha}}^{(t)}) =  \left( 
{\bf \omega} {\bf\alpha}_j^{(t)} + {\bf W}_j^{-1}({\bf\alpha}^{(t)}) {\bf U}_j({{\bf\alpha}}^{(t)}) \right)$ where ${\bf U}_j({\widehat{\bf\alpha}}^{(t)})$ is given by \eqref{u_alpha} and $\lambda_j^*$ is the tuning parameter of the $j$-th expert, $j=1,\ldots,J$.
\end{lemma}
There are two sets of tuning parameters required for the ridge estimate of the FMPRE model \eqref{fmpr}. The first set $(\lambda_1,\ldots,\lambda_J)$ was used to tune the penalty term involved in the Poisson regressions and the second set  $(\lambda_1^*,\ldots,\lambda_J^*)$ was used for the $J$ experts. 
The tuning parameters of the ridge estimation method can be estimated by various proposals. 
Here we follow \cite{liu2003using,hoerl1975ridge} and estimate the tuning parameters by
$\widehat{\lambda}_j=p/{\widehat{\bf\beta}}_{j,{\text ML}}^\top {\widehat{\bf\beta}}_{j,{\text ML}}$ and $\widehat{\lambda}_j^*=q/{\widehat{\bf\alpha}}_{j,{\text ML}}^\top {\widehat{\bf\alpha}}_{j,{\text ML}}$ where ${\widehat{\bf\alpha}}_{j,{\text ML}}$ and ${\widehat{\bf\alpha}}_{j,{\text ML}}$ for $j=1,\ldots,J$ correspond to the ML estimates of the coefficients for the component regressions and experts, respectively.  Finally, the proposed E-, S- and M-steps are alternated until the chain meets $|\ell(\bpsi^{(t+1)}|{\bf y})-\ell(\bpsi^{(t)}|{\bf y})|< \epsilon$  or reaches a pre-specified maximum number of iterations.

\subsection{Liu-type Estimation for FMPRE}\label{sub:liu} 
While the ridge estimation method, described in Subsection \ref{sub:ridge}, was proposed to deal with multicollinearity issues in the FMPRE model \eqref{fmpr}, the ridge proposals may not able to fully cope with the ill-conditioned design matrices when the multicollinearity is severe 
in the component regressions and expert classes of  \eqref{fmpr}. 
When the multicollinearity is high, a small value for ridge tuning parameter may not be adequate to handle the ill-conditioned design matrices. 
On the other side, a large value of the tuning parameter may lead to a large bias in the estimation process, affecting the performance of the ridge estimates. 
In this situation, \cite{liu2003using} proposed a shrinkage method using a new penalty to deal with the high multicollinearity problem.  
In this subsection, we develop the Liu-type (LT) estimates for the parameters of the FMPRE model \eqref{fmpr} in the presence of high multicollinearity. 

One can find the LT estimates for the parameters of FMPRE \eqref{fmpr} as a solution to the incomplete likelihood function \eqref{ll_inc} subject to the LT penalty 
where the LT penalty is given by 
\begin{align}\label{lt_penalty}
(-\frac{d}{\lambda^{1/2}}) {\widehat \bpsi} = \lambda^{1/2} \bpsi + \epsilon',
\end{align}
where $\lambda > 0$, similar to Subsection \ref{sub:ridge}, represents the ridge tuning parameter,  $d \in \R$ is the bias correction parameter and 
${\widehat \bpsi}$ denotes any estimate for the parameters of the FMPRE.  
There are various candidates for ${\widehat \bpsi}$ in \eqref{lt_penalty}. We assumed that ${\widehat \bpsi}$ is obtained from the ridge estimate ${\widehat \bpsi}_R$ throughout this manuscript. 

Like previous subsections, we shall obtain the parameters of the FMPRE model 
\eqref{fmpr} by maximum likelihood estimates using the LT penalty; however, the log-likelihood function  \eqref{ll_inc} subject to LT penalty \eqref{lt_penalty} is not tractable with respect to the coefficients of the Poisson components and the expert classes of the mixture model. 
For this reason, we introduce again the latent variables ${\bf Z}_i=(Z_{i1},\ldots, Z_{iJ})$ controlling the component memberships of the observations $i=1,\ldots,n$.
Viewing  $({\bf X},{\bf Z},{\bf y})$ as the complete data, we construct the complete likelihood function of $\bpsi$ and develop the SEM algorithm to iteratively obtain the MLE for parameters using the LT penalty.  
The E- and S-steps of the SEM algorithm remain the same as the E- and S- steps discussed in Subsection \ref{sub:ridge}. 
In the M-step, we use Lemma \ref{lem:beta_liu} to update the coefficients of the component regressions in the $(t+1)$-th iteration from the conditional expectation of the log-likelihood $Q_2({\bf \beta},{\bpsi}^{(t)})$ in \eqref{Q2} under LT penalty using partitioned data ${\bf P}^{(t+1)}$ from the S-step. 
\begin{lemma}\label{lem:beta_liu}
    Under the assumption of  FMPRE \eqref{fmpr}, the coefficients of the Poisson component regressions ${\bf\beta}_j, j=1,\ldots,J$ in the $(t+1)$-th iteration of the SEM algorithm is updated by 
       \[
    {\widehat{\bf\beta}}_{j,{\text LT}}^{(t+1)} = 
    \left( {\bf X}_j^\top {\bf W}_j {\bf X}_j + \lambda_j \I \right)^{-1}
    \left(
    {\bf X}_j^\top {\bf W}_j {\bf z}_j^* + d_j {\widehat{\bf\beta}}_{j,{R}}
    \right),
    \]
    as a solution to the maximization of the conditional expectation of log-likelihood $Q_2({\bf \beta},{\bpsi}^{(t)})$ under LT penalty, where ${\bf W}_j$ is diagonal weight matrix of size $n_j$ with $i$-th diagonal entry $\mu_j({\bf x}_i)$ and 
    \[{\bf z}^*_j = \left\{ {\bf X}_j {\widehat{\bf \beta}}_j^{(t)} +  \frac{1}{\mu_j({\bf x}_i)} 
     [{\bf y} - \mu_j({\bf x}_i)]\right\}.
     \] 
\end{lemma}
In the next step, we focus on the conditional expectation of the log-likelihood $Q_1({\bf \alpha}, {\bpsi}^{(t)})$ in \eqref{Q1} under LT penalty to update the coefficients of the expert classes. 
To do that, we use the stochastic assignments of $\Tilde{{\bf Z}}^{(t+1)}_i,i=1,\ldots,n$ 
from the S-step and rewrite the $Q_1({\bf \alpha},{\bpsi}^{(t)})$. 
Accordingly, the coefficients of expert classes can be updated based on the log-likelihood of the multinomial logit regression under the LT penalty \eqref{lt_penalty}. 
Similar to Subsection \ref{sub:ridge}, there is no closed form for the ML estimates of $\ba_j, j=1,\ldots,J$ under the LT penalty. 
Hence, we employ the coordinate descent algorithm to update  ${\widehat{\bf\alpha}}_j^{(t+1)}$ sequentially.
In each iteration of the coordinate descent algorithm, given the coefficients of the other experts, we apply the IRWLS  to obtain the coefficients of the experts ${\bf\alpha}_j, j=1,\ldots,J$, sequentially.   
Lemma \ref{lem:alpha_liu} shows how   ${\widehat{\bf\alpha}}_j^{(t+1)}$ are updated using the IRWLS method under LT penalty in the M-step of the SEM algorithm.  

\begin{lemma}\label{lem:alpha_liu}
    Under the assumption of FMPRE \eqref{fmpr}, the coefficients of the $j$-th expert, ${\bf\alpha}_j$, under LT penalty in the $(t+1)$-th iteration of the SEM is updated by
    \[
    {\widehat{\bf\alpha}}_{j,{\text LT}}^{(t+1)} = 
    \left({\bf\omega}^\top {\bf W}_j({{\bf\alpha}}^{(t)}) {\bf\omega} + \lambda^*_j \I\right)^{-1} 
    \left(
     {\bf\omega}^\top {\bf W}_j({{\bf\alpha}}^{(t)}) 
    {\bf v}_j({{\bf\alpha}}^{(t)}) + d_j^* {\widehat{\bf\alpha}}_{j,{\text R}}
    \right),
    \]
where  ${\bf W}_j({{\bf\alpha}}^{(t)})$ is computed by \eqref{w_alpha} and
${\bf v}_j({{\bf\alpha}}^{(t)}) =  \left( 
{\bf \omega} {{\bf\alpha}}_j^{(t)} + {\bf W}_j^{-1}({{\bf\alpha}}^{(t)}) {\bf U}_j({{\bf\alpha}}^{(t)}) \right)$ where ${\bf U}_j({{\bf\alpha}}^{(t)})$ is given by \eqref{u_alpha} and $\lambda_j^*$ and $d_j^*$ are the tuning parameters of the $j$-th expert, $j=1,\ldots,J$.
\end{lemma}
As one can see from Lemmas \ref{lem:beta_liu} and \ref{lem:alpha_liu}, the Liu-type method requires four sets of tuning parameters in estimating the coefficients of component regressions and the expert classes of FMPRE model \eqref{fmpr}. 
Similar to the ridge method, we require two sets of tuning parameters $\lambda_j$ and $\lambda_j^*$ for $j=1,\ldots,J$ in estimating the coefficients of the regressions and the expert classes, respectively. 
Accordingly, we follow \cite{liu2003using,ghanem2022unsupervised} and estimate 
$\widehat{\lambda}_j=p/{\widehat{\bf\beta}}_{j,{\text R}}^\top {\widehat{\bf\beta}}_{j,{\text R}}$ and $\widehat{\lambda}_j^*=q/{\widehat{\bf\alpha}}_{j,{\text R}}^\top {\widehat{\bf\alpha}}_{j,{\text R}}$ where ${\widehat{\bf\alpha}}_{j,{\text R}}$ and ${\widehat{\bf\beta}}_{j,{\text R}}$  are, respectively, the ridge estimates of ${\bf\alpha}_j$ and ${\bf\beta}_{j}$ for $j=1,\ldots,J$.
In addition to the ridge tuning parameters, the Liu-type method requires two sets of bias correction parameters $d_j$ and $d_j^*$ in estimating the parameters of the FMPRE model.  
Similar to \cite{inan2013liu,ghanem2022liu},  we develop an operational approach to estimate the optimum value of $d_j$ minimizing the mean squared error (MSE) of LT estimates ${\widehat{\bf\beta}}_{j,\text{LT}}^{(t+1)}$ using the partitions declared in the S-step.  
It is easy to show that the $\text{MSE}({\widehat{\bf\beta}}_{j,\text{LT}})$ is calculated by
 \[
 \text{MSE}({\widehat \bb}_{LT,j}) = 
 tr\left[ \text{Var}({\widehat \bb}_{LT,j}) \right] +
 || \E({\widehat \bb}_{j,LT}) -\bb_j||_2^2, 
 \]
 where
 \begin{align*}
  \text{Var}({\widehat \bb}_{LT,j}) =& 
 \left({\bf X}_j^\top {\bf W}_j {\bf X}_j + \lambda_j \I \right)^{-1}  
 \left({\bf X}_j^\top {\bf W}_j {\bf X}_j - d_j \I\right)
 \left({\bf X}_j^\top {\bf W}_j {\bf X}_j + \lambda_j \I \right)^{-1}  
 \left({\bf X}_j^\top {\bf W}_j {\bf X}_j \right)  \\
 &  
 \left({\bf X}_j^\top {\bf W}_j {\bf X}_j + \lambda_j \I \right)^{-1}
 \left({\bf X}_j^\top {\bf W}_j {\bf X}_j - d_j \I\right)
 \left({\bf X}_j^\top {\bf W}_j {\bf X}_j + \lambda_j \I \right)^{-1},
 \end{align*}
 and 
 \begin{align*}
    \E({\widehat \bb}_{LT,j}) 
 & = \left({\bf X}_j^\top {\bf W}_j {\bf X}_j + \lambda_j \I \right)^{-1}  
 \left({\bf X}_j^\top {\bf W}_j {\bf X}_j - d_j \I\right)
 \left({\bf X}_j^\top {\bf W}_j {\bf X}_j + \lambda_j \I \right)^{-1}  
 {\bf X}_j^\top {\bf W}_j {\mu}_j({\bf X}_j).
 \end{align*}
We must also develop an operational approach to estimate the bias correction parameters $d_j^*$ in each expert class $j=1,\ldots,J$. 
Like the Poisson component regressions, we estimate the $d_j^*$ by 
minimizing the $\text{MSE}({\widehat{\bf\alpha}}_{j,\text{LT}})$. One can easily derive
 \[
 \text{MSE}({\widehat \ba}_{LT,j}) = 
 tr\left[ \text{Var}({\widehat \ba}_{LT,j}) \right] +
 || \E({\widehat \ba}_{j,LT}) -\ba_j||_2^2, 
 \]
 where
 \begin{align*}
  \text{Var}({\widehat \ba}_{LT,j}) =& 
 \left({\bf \omega}^\top {\bf W}_j({\bf\alpha}) {\bf \omega} 
 + \lambda_j^* \I \right)^{-1}  
 \left({\bf \omega}^\top {\bf W}_j({\bf\alpha}) {\bf \omega} - d_j^* \I\right)
 \left({\bf \omega}^\top {\bf W}_j({\bf\alpha}) {\bf \omega} + \lambda_j^* \I \right)^{-1}  \\
 &\left({\bf \omega}^\top {\bf W}_j({\bf\alpha}) {\bf \omega} \right)  
 \left({\bf \omega}^\top {\bf W}_j({\bf\alpha}) {\bf \omega} + \lambda_j^* \I \right)^{-1}
 \left({\bf \omega}^\top {\bf W}_j({\bf\alpha}) {\bf \omega} - d_j^* \I\right)
 \left({\bf \omega}^\top {\bf W}_j({\bf\alpha}) {\bf \omega} + \lambda_j^* \I \right)^{-1},
 \end{align*}
 and 
 \begin{align*}
    \E({\widehat \ba}_{LT,j}) 
 & = \left({\bf \omega}^\top {\bf W}_j({\bf\alpha}) {\bf \omega} 
 + \lambda_j^* \I \right)^{-1}  
 \left({\bf \omega}^\top {\bf W}_j({\bf\alpha}) {\bf \omega} - d_j^* \I\right)
 \left({\bf \omega}^\top {\bf W}_j({\bf\alpha}) {\bf \omega}
 + \lambda_j^* \I \right)^{-1}  
 {\bf \omega}^\top {\bf W}_j({\bf\alpha}) \pi_j({\bf \omega},{\bf \alpha}).
 \end{align*}
As both $\text{MSE}({\widehat{\bf\beta}}_{j,\text{LT}})$ and $\text{MSE}({\widehat{\bf\alpha}}_{j,{\text LT}})$ depend on the true values of the coefficients ${\bb_j}$ and $\ba_j$, following \cite{inan2013liu,ghanem2022liu}, we use ${\widehat{\bb}}_{j,{R}}$ and ${\widehat{\ba}}_{j,{R}}$ as the true values of the coefficients in estimating the bias correction coefficients  $(d_j,d_j^*)$ for $j=1,\ldots,J$. 
Finally, the E-, S- and M-steps of the SEM algorithm under LT estimation method are alternated until the stopping rule $|\ell(\bpsi^{(t+1)}|{\bf y})-\ell(\bpsi^{(t)}|{\bf y})|< \epsilon$ satisfies or the algorithm reaches a pre-specified maximum number of iterations.

\section{Simulation Studies}\label{sec:sim}
In this section, we compare the performance of the ML, Ridge and Liu-type methods in estimating the parameters of the FMPRE model \eqref{fmpr} in multicollinearity. To do that, we present two simulation studies to evaluate the estimation performance of the proposed methods and investigate the impact of sample size, multicollinearity levels and the number of the components of mixture models on the parameter estimates. Accordingly, in the first simulation study, we focus on the FMPRE model with two Poisson component regressions, where each component regression and the expert class consist of four correlated covariates. In contrast, in the second study, the underlying FMPRE comprises three components, each with two correlated covariates.  

\begin{table}[H]
\caption{\small{The median (M), lower (L) and upper (U) bounds of 90\% CIs for ${\sqrt{\text{MSE}}}$ and Accuracy of the methods in estimation and classification of the FMPRE model with two component regressions when $\rho=0.85$ and $n = 100$.}}
\label{2m_n100_r0.85}
\begin{centering}
\begin{tabular}{ccccccccccc}
\hline 
 &  & \multicolumn{2}{c}{} & \multicolumn{3}{c}{${\sqrt{\text{MSE}}}$} &  & \multicolumn{3}{c}{Accuracy}\tabularnewline
\cline{5-7} \cline{9-11} 
$\phi$ & EM & $\ensuremath{\ensuremath{\boldsymbol{\Psi}}}$ &  & M & L & U &  & M  & L & U\tabularnewline
\hline 
0.90 & LS & ${\bb}$ &  & 2.31 & 0.02 & 231.9 &  & 0.50 & 0.17 & 0.93\tabularnewline
\cline{5-7} 
 &  & ${\ba}$ &  & 0.224 & 0.224 & 53.3 &  &  &  & \tabularnewline
\cline{2-11} 
 & Ridge & ${\bb}$ &  & 0.702  & 0.011 & 44.6 &  & 0.70 & 0.17 & 0.98\tabularnewline
\cline{5-7} 
 &  & ${\ba}$ &  & 0.224 & 0.218 & 28.7 &  &  &  & \tabularnewline
\cline{2-11} 
 & LT & ${\bb}$ &  & 0.163 & 0.011 & 12.05 &  & 0.88 & 0.17 & 0.98\tabularnewline
\cline{5-7} 
 &  & ${\ba}$ &  & .0.268 & 0.224 & 6.4 &  &  &  & \tabularnewline
\hline 
0.95 & LS & ${\bb}$ &  & 2.83 & 0.025 & 218.9 &  & 0.56 & 0.26 & 0.94\tabularnewline
\cline{5-7} 
 &  & ${\ba}$ &  & 0.224  & 0.224 & 63.8 &  &  &  & \tabularnewline
\cline{2-11} 
 & Ridge & ${\bb}$ &  & 0.846  & 0.014 & 38.4 &  & 0.72 & 0.26 & 0.98\tabularnewline
\cline{5-7} 
 &  & ${\ba}$ &  & 0.224 & 0.218 & 33.2 &  &  &  & \tabularnewline
\cline{2-11} 
 & LT & ${\bb}$ &  & 0.129 & 0.012 & 35.3 &  & 0.91 & 0.26 & 0.99\tabularnewline
\cline{5-7} 
 &  & ${\ba}$ &  & 0.302 & 0.224 & 6.4 &  &  &  & \tabularnewline
\hline 
\end{tabular}
\par\end{centering}
\end{table}

In the first simulation study, the underlying FMPRE comprises two Poisson components where we generate correlated covariates $x_1,\ldots,x_4$ and $\omega_1\ldots,\omega_4$ to simulate the multicollinearity in the Poisson regressions and the expert classes. Here, we follow \cite{inan2013liu,ghanem2022liu} to introduce multicollinearity in both sets of covariates. To do so, we employed two parameters $\phi$ and $\rho$ to generate multicollinearity in the design matrices where $\phi^2$ denote the correlation levels between the first two covariates and $\rho^2$  the correlation between the last two covariates. We first generate random samples $u_{ij}, i=1,\ldots,n;j=1,\ldots,5$ from a standard normal distribution to simulate the correlated predictors. The correlated predictors are then simulated by
\[
x_{i,l_1} = (1-\phi^2) u_{i,l_1} + \phi u_{i,5}, ~~~ l_1=1,2,
\]
\[
x_{i,l_2} = (1-\rho^2) u_{i,l_2} + \rho u_{i,5}, ~~~ l_2=3,4.
\]   
In a similar vein, we generate the correlated covariates in the multinomial regressions of the expert class such that $\phi^2=cor(\omega_1,\omega_2)$ and $\rho^2=cor(\omega_3,\omega_4)$.
To study the effects of colinearity on the estimates, we consider four sets of correlations $(\phi,\rho)=\{(0.85,0.90),(0.85,0.95), (0.90,0.90),(0.90,0.95)\}$ to simulate the multicollinearity levels in the Poisson regressions as well as the expert classes. Subsequently, we create the count responses from the FMPRE model \eqref{fmpr} using the generated predictors and the concomitants. The coefficients of the model were set to ${\bpsi}=(\ba_{1},\bb_{1},\bb_{2})$ where, $\bb_{1} = (1,1,2,3,0.5)$, $\bb_{2} = (-1,-1,-2,-0.5,-2)$, $\ba_{1} = (0.5,-1,-1,0.3,-3)$ where the second expert class was set as the reference.  

\begin{table}[H]
\caption{\small{The median (M), lower (L) and upper (U) bounds of 90\% CIs for ${\sqrt{\text{MSE}}}$ and Accuracy of the methods in estimation and classification of the FMPRE model with two component regressions when $\rho=0.90$ and $n = 100$.}}
\begin{centering}
\label{2m_n100_r0.90}
\begin{tabular}{ccccccccccc}
\hline 
 &  & \multicolumn{2}{c}{} & \multicolumn{3}{c}{${\sqrt{\text{MSE}}}$} &  & \multicolumn{3}{c}{Accuracy}\tabularnewline
\cline{5-7} \cline{9-11} 
$\phi$ & EM & $\ensuremath{\ensuremath{\boldsymbol{\Psi}}}$ &  & M & L & U &  & M  & L & U\tabularnewline
\hline 
0.90 & LS & ${\bb}$ &  & 2.558 & 0.024 & 231.8 &  & 0.58 & 0.22 & 0.92\tabularnewline
\cline{5-7} 
 &  & ${\ba}$ &  & 0.224 & 0.224 & 54.1 &  &  &  & \tabularnewline
\cline{2-11} 
 & Ridge & ${\bb}$ &  & 0.720 & 0.012 & 37.4 &  & 0.74 & 0.22 & 0.98\tabularnewline
\cline{5-7} 
 &  & ${\ba}$ &  & 0.224 & 0.224 & 27.0 &  &  &  & \tabularnewline
\cline{2-11} 
 & LT & ${\bb}$ &  & 0.205 & 0.012 & 16.0 &  & 0.84 & 0.22 & 0.99\tabularnewline
\cline{5-7} 
 &  & ${\ba}$ &  & 0.273 & 0.224 & 5.4 &  &  &  & \tabularnewline
\hline 
0.95 & LS & ${\bb}$ &  & 3.338 & 0.029 & 275.9 &  & 0.52 & 0.23 & 0.94\tabularnewline
\cline{5-7} 
 &  & ${\ba}$ &  & 0.224 & 0.224 & 61.3 &  &  &  & \tabularnewline
\cline{2-11} 
 & Ridge & ${\bb}$ &  & 0.860 & 0.016 & 64.4 &  & 0.75 & 0.23 & 0.99\tabularnewline
\cline{5-7} 
 &  & ${\ba}$ &  & 0.224 & 0.224 & 32.4 &  &  &  & \tabularnewline
\cline{2-11} 
 & LT & ${\bb}$ &  & 0.148 & 0.014 & 94.0 &  & 0.91 & 0.23 & 0.98\tabularnewline
\cline{5-7} 
 &  & ${\ba}$ &  & 0.280 & 0.224 & 7.0 &  &  &  & \tabularnewline
\hline 
\end{tabular}
\par\end{centering}
\end{table}

As described above, we collected a training sample of size $n=\{100,200\}$  from the FMPRE population. We estimated the population parameters using ML, Ridge and Liu-type methods, as discussed in Section \ref{sec:stat}. We then assessed the estimation performance of the developed methods by calculating the squared root of the mean squared errors in estimating the coefficients of the Poisson regressions 
as $\sqrt{\text{MSE}(\widehat \bb)} = \left( \sum_{j=1}^{J} 
({\widehat\bb}_j -\bb)^\top({\widehat\bb}_j -\bb)/n \right)^{1/2}$ and the coefficients of the expert classes 
$\sqrt{\text{MSE}(\widehat \ba)}= \left( \sum_{j=1}^{J} 
({\widehat\ba}_j -\ba)^\top({\widehat\ba}_j -\ba)/n \right)^{1/2}$
where $\bb$ and $\ba$ represent the true values for parameters of the underlying population. 
In the case of the classification, once the parameters of the FMPRE were estimated, we used the trained models to obtain the classification performance of ML, Ridge and Liu-type methods in predicting the component membership of a validation data set. 
The validation data set of size $n_t=100$ was selected from the underlying mixture model independent from the training data sets. 

We computed the classification performance of the developed methods using the measure of accuracy, (TP+TN)/N, where TP denotes the true position case where the model classified correctly an individual to the first component and TN stands for the true negative case where the model classified correctly an individual to the second component of the population. To take sampling variability into account,  we replicated the entire data generation, estimation, and classification 2000 times utilizing the ML, Ridge, and LT methods.  We then computed the median (50\%) and 90\% intervals for  $\sqrt{\text{MSE}}$ and accuracy measures to report the estimation and classification performances of the methods over 2000 replicates. To compute the 90\%  interval for the performance measure, we first sorted the 2000 replicates. Then, we reported the 5 and 95 percentiles, respectively,  as the lower and upper bounds of the interval for the corresponding measure. 

\begin{table}[H]
\caption{\small{The median (M), lower (L) and upper (U) bounds of 90\% CIs for ${\sqrt{\text{MSE}}}$ and Accuracy of the methods in estimation and classification of the FMPRE model with three component regressions when $\rho=0.90$ and $n = 300$.}}
\begin{centering}
\label{3m_n300_r0.90}
\begin{tabular}{cccccccccc}
\hline 
 & \multicolumn{2}{c}{} & \multicolumn{3}{c}{${\sqrt{\text{MSE}}}$} &  & \multicolumn{3}{c}{Accuracy}\tabularnewline
\cline{4-6} \cline{8-10} 
EM & $\ensuremath{\ensuremath{\boldsymbol{\Psi}}}$ &  & M & L & U &  & M  & L & U\tabularnewline
\hline 
LS & ${\bb}$ &  & 0.44 & 0.25 & 5.86 &  & 0.42 & 0.15 & 0.81\tabularnewline
\cline{2-6} 
 & ${\ba}$ &  & 0.97 & 0.30 & 11.79 &  &  &  & \tabularnewline
\hline 
Ridge & ${\bb}$ &  & 0.37 & 0.18 & 2.26 &  & 0.42 & 0.14 & 0.86\tabularnewline
\cline{2-6} 
 & ${\ba}$ &  & 0.57 & 0.19 & 3.13 &  &  &  & \tabularnewline
\hline 
LT & ${\bb}$ &  & 0.33 & 0.17 & 0.78 &  & 0.40 & 0.09 & 0.91\tabularnewline
\cline{2-6} 
 & ${\ba}$ &  & 0.25 & 0.09 & 3.02 &  &  &  & \tabularnewline
\hline 
\end{tabular}
\par\end{centering}
\end{table}

Tables \ref{2m_n100_r0.85}, \ref{2m_n100_r0.90} and \ref{2m_n200_r0.85}-\ref{2m_n200_r0.90} demonstrate the results of the first simulation study. One observes that the ML method leads to significantly unreliable results in estimating the coefficients of the Poisson component regressions and the expert classes.  Interestingly, we observe that the estimation performance of the ML method deteriorates further as the multicollinearity level increases. While ML estimates appear unreliable, their shrinkage counterparts, like ridge and LT methods, on average, result in more reliable outcomes in estimating the parameters of the FMRPE, even in the presence of high multicollinearity. 
For example, 90\% interval for  $\sqrt{\text{MSE}(\widehat \bb)}$  under the ML method is given by (0.02, 231.9) when $(\phi,\rho)=(0.9,0.85)$ in Table \ref{2m_n100_r0.85}, while the ridge method gives (0.011, 44.5) and this value reduces to (0.21, 12.05) for the Liu-type method. A similar pattern, on average, is seen in estimating the unknown coefficients of the expert classes, where the shrinkage estimates outperform their ML method in the presence of multicollinearity. Comparing the shrinkage proposals, we see that the Liu-type proposals $({\widehat\bb}_\text{LT}, {\widehat\ba}_\text{LT})$, on average, perform better than their ridge counterparts in estimating the parameters of the FMPRE even in the presence of server multicollinearity issue. In the case of classification performance, it seems that the accuracy across all three methods, on average, is almost the same. This is compatible with the literature's findings of \cite{inan2013liu} and \cite{ghanem2022liu}.

The second simulation examines the performance of the proposed methods in estimating the parameters of the FMPRE model consisting of three Poisson regression models. In this study, we set two covariates for the Poisson component regressions and the expert classes. We assume that the correlation level between the two covariates is $\rho=cor(x_1,x_2)=cor(\omega_1,\omega_2)$ where we set $\rho=\{0.9,0.95\}$. Similar to the first study, we generated the correlated covariates and the Poisson responses from FMPRE with the true parameters $\bb_1=(0.85,-1,2)$, 
$\bb_2=(1,0.5,1)$,
 $\bb_3=(-2,2,-2)$,
 $\ba_{1} = (0.5,-1,-1)$ and $\ba_{2} = (0.1,1,0.05)$ where the third class was treated as the reference. 
  Similar to the first simulation study, we collected the sample of size $n=300$ for the underlying population and estimated the parameters of the underlying population using ML, Ridge and Liu-type methods as described in Section \ref{sec:stat}. 
  The results of this study are presented in Tables \ref{3m_n300_r0.90} and \ref{3m_n300_r0.95}. Although the classification performances under the three methods, on average, are the same, the shrinkage methods perform better than the ML method, leading to more reliable results in estimating the coefficients of the Poisson regressions and expert classes of the FMPRE. We also see that Liu-type shrinkage estimates, on average, outperform their Ridge counterparts in estimating both the parameters of the Poisson component regressions and the expert classes.

\section{Real Data Analysis}\label{sec:real}
Cardiovascular diseases (CVDs) remain one of the dominant causes of morbidity, disability and mortality in western communities. 
CVDs lead to various heart diseases. Heart problems include coronary heart disease, difficulties with the heart valves, heart failure, and arrhythmia, to name a few. Heart diseases impose significant medical and financial pressure on the healthcare systems. For example, in the US, an average heart attack occurs every 40 seconds. About 1 out of 5 cardiac disorders are silent; the patient is unaware of the condition as it progresses. The annual cost of heart disease, including medicine, care services and death, accounted for 
\$239.9 billion in 2018 only in the US \citep{tsao2023heart,pearson2016modeling,zeiger2012evaluation}. 
Hence, it is crucial to develop statistical methods to more efficiently and reliably evaluate the impact of the predictors on heart disease.  
These methods enable national health systems to monitor the prevalence of heart diseases to plan better the well-being of the communities.  

The QRS complex in an electrocardiogram (ECG) comprises three successive Q, R, and S waves. Using these waveforms as a basis, the complex demonstrates de-polarization of the heart ventricles, which collect and evacuate blood into the lung tissues and other peripheral areas of the body.  The T wave in an ECG, on the other hand, indicates that the ventricles have repolarized. 
In an ECG, the time between the conclusion of the QRS complex and the start of the T wave is represented by the ST segment index. 
The ST segment shows a modest upward slope for a healthy heart; however, a depressed and downward slope may signify a heart condition such as coronary ischemia. The relative ST segment depression and ST segment slop as two ECG criteria are typically used as predictors for diagnosing heart disease \citep{lanza2004diagnostic,baker2019exercise,ryu2019risk}.  

This data analysis focused on the Cleveland Clinic Foundation heart disease data set, publicly available at the University of California Irvine Machine Learning repository.  This data set contains 303 observations with 76 attributes; however, the previous research studies typically worked with a subset of 14 attributes \citep{taneja2013heart}. 
 According to the mixed type of heart data set, \cite{mufudza2016poisson} proposed a mixture of Poisson regression models for the classification of instances using the heart rate stage where the response variable was the count of the rate at which heart disease was diagnosed.  The zero value in response represents no disease, while values 1, 2, 3 and 4 indicate different levels of heart disease. 

\begin{table}[H]
\caption{\small{The median (M), lower (L) and upper (U) bounds of 90\% CIs for ${\sqrt{\text{MSE}}}$ and Accuracy of the methods in estimation and classification of the heart disease population when $n = 30$.}}
\label{tab:real_30}
\begin{centering}
\begin{tabular}{cccccccccc}
\hline 
 & \multicolumn{2}{c}{} & \multicolumn{3}{c}{${\sqrt{\text{MSE}}}$} &  & \multicolumn{3}{c}{Accuracy}\tabularnewline
\cline{4-6} \cline{8-10} 
EM & $\ensuremath{\ensuremath{\boldsymbol{\Psi}}}$ &  & M & L & U &  & M  & L & U\tabularnewline
\hline 
LS & ${\bb}$ &  & 0.22 & 0.086 & 14.40 &  & 0.92 & 0.85 & 0.99\tabularnewline
\cline{2-6} 
 & ${\ba}$ &  & 0.42 & 0.083 & 6.70 &  & . &  & \tabularnewline
\hline 
Ridge & ${\bb}$ &  & 0.15 & 0.065 & 1.90 &  & 0.93 & 0.87 & 0.99\tabularnewline
\cline{2-6} 
 & ${\ba}$ &  & 0.26 & 0.054 & 1.81 &  &  &  & \tabularnewline
\hline 
LT & ${\bb}$ &  & 0.12 & 0.042 & 0.44 &  & 0.94 & 0.88 & 1.00\tabularnewline
\cline{2-6} 
 & ${\ba}$ &  & 0.15 & 0.058 & 0.46 &  &  &  & \tabularnewline
\hline 
\end{tabular}
\par\end{centering}
\end{table}

Following \cite{mufudza2016poisson}, we applied the mixture of the Poisson regression models with an expert to the data set where the heart disease rate stage was treated again as the count response variable of the  FMPRE model \eqref{fmpr}. We also treated ST depression $Z_1$ and ST slope $Z_2$ criteria of the ECG as two covariates of the regression components and the expert classes of the FMPRE model \eqref{fmpr}.
The correlation between the two covariates is $\rho=cor(Z_1, Z_2)=0.58$, indicating the  multicollinearity in the FMPRE model. 
We applied the ML method for a mixture of the Poisson regression models to the entire population of 297 instances (after removing the missing data); the Bayesian information criteria (BIC) suggests that the mixture of two Poisson regression models was the best fit to the data set \citep{mufudza2016poisson}.  
We then treated the ML estimates using the entire heart population as the true values of the parameters of the FMPRE. 
Similar to Section \ref{sec:sim},  we applied ML, Ridge and Liu-type methods in estimating the parameters of the underlying FMPRE using training data of size $n=\{30,50\}$ and test sample of size 100 taken independently from the training phase. 
We replicated the data collection and estimation method 2000 times and calculated the estimation $\sqrt{\text{MSE}}$'s and classification accuracy as described in Section \ref{sec:sim}. 

Tables \ref{tab:real_30} and \ref{tab:real_50} provide the results of the heart disease data analysis. In multicollinearity, the results are highly skewed. For this reason, we reported the median (M) and 90\% interval for the estimation and classification measures.  
 We computed the lower (L) and upper  (U) bounds of the 90\%  intervals by 5 and 90 percentiles from 2000 replicates of the estimation and classification measures. Similar to the results of the simulation studies, from Tables \ref{tab:real_30} and \ref{tab:real_50}, we observe that the classification accuracy of all the methods is almost the same. Despite this, the ML method leads to larger variability, on average, in both the median and the length of the interval in estimating the coefficients of the Poisson regression components and the experts.  
Unlike ML estimates, the Ridge and Liu-type shrinkage methods could handle the multicollinearity and result in more reliable estimates for coefficients of the FMPRE model. 
 Comparing the Liu-type proposal with its Ridge counterpart, we observe that ${\widehat \bb}_{\text{LT}}$ and ${\widehat \ba}_{\text{LT}}$, on average, almost always outperform their ridge counterparts in both the median and the length of the intervals for estimation errors. 
 Consequently, we recommend that practitioners use the LT  method in estimating the parameters of the heart disease population using the FMPRE model in the presence of multicollinearity.


\section{Summary and Concluding Remarks}\label{sec:con}

Count data types, such as heart disease research, are widely used in medical surveys. The mixture of Poisson regression models is a practical and robust data analysis tool to incorporate the information in a set of covariates, possibly of mixed types, to explain the count response variable in heterogeneous populations. The maximum likelihood method is considered a common approach in estimating the parameters of the mixture of Poisson regression models. The ML estimates are highly affected when there is a collinearity issue between the covariates of the regression models, such that the method may provide unreliable estimates for the coefficients of the model.  

In this research, we developed Ridge and Liu-type shrinkage estimation methods to deal with the multicollinearity in the finite mixture of Poisson regression models with experts (FMPRE). Through various simulation studies, we compared the performance of the developed shrinkage estimates with their ML counterparts. Numerical studies demonstrated that the classification performance of the three methods is almost the same; however,  the ridge and Liu-type shrinkage methods result in more reliable estimates for the coefficients of the regression components and the expert classes. We also observed that the Liu-type estimates, on average, almost consistently outperformed the Ridge method in estimating the parameters of the FMPRE. Finally, we applied the developed methods to the Cleveland Clinic Foundation heart data set to analyze the heart disease rate stages.

\section*{Acknowledgment}  
Armin Hatefi acknowledges the research support of the Natural Sciences and Engineering Research Council of Canada (NSERC).  
The authors thank the UCI machine learning repository for the valuable assistance received by using the repository.

\nocite{*}
\bibliographystyle{plainnat}
{\small\bibliography{LT-reg-bib}
}

\newpage
\section{Appendix} 

\subsection{Proof of Lemma \ref{lem:beta_ridge}}
From the conditional expectation of the log-likelihood \eqref{Q2} and
the partition ${\bf P}^{(t+1)}$ from S-step, the ${\widehat{\bf\beta}_{j,R}}$ in the $(t+1)$-th iteration is updated as solution to 
\begin{align}
    Q_2^R({\bf \beta}_j,{\bpsi}^{(t)})=
    \sum_{i=1}^{n_j} 
    \log\left\{ y_i \log(\mu_j({\bf x}_i)) - \mu_j({\bf x}_i) \right\} 
    - \lambda_j {\bf \beta}_j^\top
    {\bf \beta}_j/2
\end{align}
where $n_j = \sum_{i=1}^{n} \I(\Tilde{{Z}}^{(t+1)}_{ij}=1)$.
From the canonical form of the Poisson distribution, the gradient under the ridge penalty is computed by 
 \begin{align}\label{grad_ridge} \nonumber
\nabla_{\bb_j} {\bf Q}^{R}_2(\bb_j,\bpsi^{(t)}) &= 
\sum_{i=1}^{n_j}  \frac{\partial}{\partial \theta_j} \big{\{} y_i \theta_{ij} -b(\theta_{ij}) - \log(y_i!)\big{\}} \frac{\partial \theta_{ij}}{\partial b(\theta_{ij})}  \frac{\partial b(\theta_{ij})}{\partial \bb_j} - \lambda_j {\bf \beta}_j \\
&= {\bf X}_j^\top \left[ {\bf y}_j - \mu_j({\bf x}_i) \right] - \lambda_j \bb_j, 
\end{align}
where $\theta_{ij} = {\bf x}_i^\top \bb_j$ and $b(\theta_{ij}) = e^{\theta_{ij}}$. Accordingly, the Hessian matrix is given by 
\begin{align}\label{hess_ridge}
{\bf H}_{\bb_j}\left({\bf Q}^{R}_2(\bb_j,\bpsi^{(t)})\right) = - {\bf X}_j^\top {\bf W}_j {\bf X}_j - \lambda_j \I.
\end{align}
Let ${\bf U}_j= {\bf H}_{\bb_j}\left({\bf Q}^{R}_2(\bb_j,\bpsi^{(t)})\right) 
|_{\bb_j = {\widehat\bb}_j^{(t)}}$. From \eqref{grad_ridge} and \eqref{hess_ridge}, the IRWLS to update ${\widehat\bb}_j^{(t+1)}$ is given by
\begin{align*}
{\widehat\bb}_j^{(t+1)} &= 
{\bb}_j^{(t)} - {\bf H}_{\bb_j}^{-1}\left({\bf Q}^{R}_2(\bb_j,{\bpsi}^{(t)})\right) 
{\nabla_{\bb_j} {\bf Q}^{R}_2(\bb_j,{\bpsi}^{(t)})} |_{\bb_j = {\bb}_j^{(t)}}\\
& = {\bb}_j^{(t)} + {\bf U}_j^{-1} 
\left\{  {\bf X}_j^\top
\left[ {\bf y}_j - {\mu}_j({\bf x}_i)\right]  - \lambda_j {\bb}_j^{(t)} \right\} \\
& = {\bf U}_j^{-1} {\bf X}_j^\top {\bf W}_j
\left\{{\bf X}_j  {\bb}_j^{(t)} + {\bf W}^{-1}_j 
\left[ {\bf y}_j - {\mu}_j({\bf x}_i)\right] \right\}\\
& = \left( {\bf X}_j^\top {\bf W}_j {\bf X}_j + \lambda_j \I\right)^{-1} {\bf X}_j^\top {\bf W}_j {\bf z}_j^*.
\end{align*}

\subsection{Proof of Lemma \ref{lem:alpha_ridge}}
The IRWLS is applied to each class of the multinomial logit regression separately in each iteration of the coordinate descent algorithm. 
For this reason,  treating all other coordinates ${\bf\alpha}_1,\ldots,{\bf\alpha}_{j-1},{\bf\alpha}_{j+1},\ldots, {\bf\alpha}_J$ fixed and from \eqref{Q1} and the partition ${\bf P}^{(t+1)}$ from S-step, we update ${\widehat{\bf\alpha}}_j^{(t+1)}$ as a solution to 
\begin{align}\label{ll_alpha_ridge} 
        Q_1^R({\bf \alpha}_j,{\bpsi}^{(t)})=
        \sum_{i=1}^{n} \I(\Tilde{{Z}}^{(t+1)}_{ij}=1) 
        \log\left\{\pi_j({\bf \omega}_i,{\bf \alpha}) \right\} - \lambda_j^* {\bf \alpha}_j^\top {\bf\alpha}_j.
\end{align}
One can easily calculate the gradient of \eqref{ll_alpha_ridge} by
 \begin{align}\label{grad_ridge_alpha} \nonumber
\nabla_{\ba_j} {\bf Q}^{R}_1(\ba_j,\bpsi^{(t)}) &= 
\sum_{i=1}^{n} 
\left(
\I(\Tilde{{Z}}^{(t+1)}_{ij}=1) - \pi_j({\bf \omega}_i,{\bf \alpha}) 
\right) {\bf \omega}_i  - \lambda_j^* {\bf\alpha}_j \\
&= {\bf \omega}^\top {\bf U}_j({\bf \alpha}) - \lambda_j^* {\bf\alpha}_j,
\end{align}
where ${\bf U}_j({\bf \alpha})$ is given by \eqref{u_alpha}. 
Accordingly, the Hessian matrix can be computed by
\begin{align}\label{hess_ridge_alpha_vec} \nonumber 
\frac{\partial^2}{\partial \ba_j \ba_l}{\bf Q}^{R}_1(\ba_j,\bpsi^{(t)})  &= - \sum_{i=1}^{n} \pi_j({\bf \omega}_i,{\bf \alpha})  
\left[ \I(\Tilde{{Z}}^{(t+1)}_{ij}=1) - \pi_j({\bf \omega}_i,{\bf \alpha})  \right] {\bf \omega}_i {\bf \omega}_i^\top \\
&= - {\bf \omega}^\top {\bf W}_j({\widehat{\bf\alpha}}) {\bf \omega} - \lambda_j \I.
\end{align}
where ${\bf W}_j({\bf\alpha}^{(t)})$ comes from \eqref{w_alpha}. 
Let ${\bf V}_j = {\bf\omega}^\top {\bf W}_j({{\bf\alpha}}^{(t)}) {\bf\omega} + \lambda^*_j \I$.  
From \eqref{grad_ridge_alpha} and \eqref{hess_ridge_alpha_vec}, the IRWLS to update ${\widehat\ba}_j^{(t+1)}$ is given by
\begin{align*}
{\widehat\ba}_j^{(t+1)} &= 
{\ba}_j^{(t)} - \frac{\partial^2}{\partial \ba_j \ba_l}{\bf Q}^{R}_1(\ba_j,\bpsi^{(t)})  
{\nabla_{\ba_j} {\bf Q}^{R}_2(\bb_j,{\widehat\bpsi}^{(l)})} 
|_{\ba_j = {\ba}_j^{(t)}}\\
& = {\ba}_j^{(t)} +  {\bf V}_j^{-1} 
\left\{
{\bf \omega}^\top {\bf U}_j({\ba}^{(t)}) - \lambda^*_j {\ba}_j^{(t)}
\right\} \\
& = {\bf V}_j^{-1} {\bf \omega}^\top {\bf W}_j({{\bf\alpha}}^{(t)})
\left\{{\bf \omega}  {\ba}_j^{(t)} + [{\bf W}_j({{\bf\alpha}}_j^{(t)})]^{-1} {\bf U}_j({\ba}_j^{(t)}) \right\}\\
& = \left( {\bf \omega}^\top {\bf W}_j({{\bf\alpha}}^{(t)})  {\bf \omega} + \lambda^*_j \I\right)^{-1} {\bf \omega}^\top {\bf W}_j({{\bf\alpha}}^{(t)})
{\bf v}_j({{\bf\alpha}}^{(t)}).
\end{align*}

\subsection{Proof of Lemma \ref{lem:beta_liu}}
From the conditional expectation of the log-likelihood \eqref{Q2} and
the partition ${\bf P}^{(t+1)}$ from S-step, the ${\widehat{\bf\beta}_{j,\text{LT}}}$ in the $(t+1)$-th iteration is updated as solution to 
\begin{align*} 
    Q_2^L({\bf \beta}_j,{\bpsi}^{(t)}) =&
    \sum_{i=1}^{n_j} 
    \log\left\{ y_i \log(\mu_j({\bf x}_i)) - \mu_j({\bf x}_i) \right\} \\
    & - \frac{1}{2} 
    \left[ \left( -\frac{d_j}{\lambda_j^{1/2}} \right) 
    {\widehat{\bf\beta}_{j,R}} - \lambda_j^{1/2} \bb_j
    \right]^\top
    \left[ \left( -\frac{d_j}{\lambda_j^{1/2}} \right) 
    {\widehat{\bf\beta}_{j,R}} - \lambda_j^{1/2} \bb_j
    \right],
\end{align*}
where $n_j = \sum_{i=1}^{n} \I(\Tilde{{Z}}^{(t+1)}_{ij}=1)$.
From the canonical form of the Poisson distribution, the gradient under the ridge penalty is computed by 
 \begin{align}\label{grad_liu} \nonumber
\nabla_{\bb_j} {\bf Q}^{L}_2(\bb_j,\bpsi^{(t)}) &= 
\sum_{i=1}^{n_j}  \frac{\partial}{\partial \theta_j} \big{\{} y_i \theta_{ij} -b(\theta_{ij}) - \log(y_i!)\big{\}} \frac{\partial \theta_{ij}}{\partial b(\theta_{ij})}  \frac{\partial b(\theta_{ij})}{\partial \bb_j} - \lambda_j {\bf \beta}_j - d_j {\widehat{\bf\beta}_{j,R}} \\
&= {\bf X}_j^\top \left[ {\bf y}_j - \mu_j({\bf x}_i) \right] 
- \lambda_j \bb_j - d_j {\widehat{\bf\beta}_{j,R}}, 
\end{align}
where $\theta_{ij} = {\bf x}_i^\top \bb_j$ and $b(\theta_{ij}) = e^{\theta_{ij}}$. Accordingly, the Hessian matrix is given by 
\begin{align}\label{hess_liu}
{\bf H}_{\bb_j}\left({\bf Q}^{L}_2(\bb_j,\bpsi^{(t)})\right) = - {\bf X}_j^\top {\bf W}_j {\bf X}_j - \lambda_j \I.
\end{align}
Let ${\bf U}_j= {\bf H}_{\bb_j}\left({\bf Q}^{L}_2(\bb_j,\bpsi^{(t)})\right) 
|_{\bb_j = {\widehat\bb}_j^{(t)}}$. From \eqref{grad_liu} and \eqref{hess_liu}, the IRWLS to update ${\widehat\bb}_j^{(t+1)}$ is given by
\begin{align*}
{\widehat\bb}_j^{(t+1)} &= 
{\bb}_j^{(t)} - {\bf H}_{\bb_j}^{-1}\left({\bf Q}^{L}_2(\bb_j,{\bpsi}^{(t)})\right) 
{\nabla_{\bb_j} {\bf Q}^{R}_2(\bb_j,{\bpsi}^{(t)})} |_{\bb_j = {\bb}_j^{(t)}}\\
& = {\bb}_j^{(t)} + {\bf U}_j^{-1} 
\left\{  {\bf X}_j^\top
\left[ {\bf y}_j - {\mu}_j({\bf x}_i)\right]  
- \lambda_j {\bb}_j^{(t)} - d_j {\widehat{\bf\beta}_{j,R}} 
\right\} \\
& = {\bf U}_j^{-1} {\bf X}_j^\top {\bf W}_j
\left\{{\bf X}_j  {\bb}_j^{(t)} + {\bf W}^{-1}_j 
\left[ {\bf y}_j - {\mu}_j({\bf x}_i)\right] \right\} - d_j  {\bf U}_j^{-1}{\widehat{\bf\beta}_{j,R}}  \\
& = \left( {\bf X}_j^\top {\bf W}_j {\bf X}_j + \lambda_j \I\right)^{-1} 
\left({\bf X}_j^\top {\bf W}_j {\bf z}_j^* -
d_j {\widehat{\bf\beta}_{j,R}}
\right)  
\end{align*}

\subsection{Proof of Lemma \ref{lem:alpha_liu}}
The IRWLS is applied to each class of the multinomial logit regression separately in each iteration of the coordinate descent algorithm. 
For this reason,  treating all other coordinates ${\bf\alpha}_1,\ldots,{\bf\alpha}_{j-1},{\bf\alpha}_{j+1},\ldots, {\bf\alpha}_J$ fixed and from \eqref{Q1} and the partition ${\bf P}^{(t+1)}$ from S-step, we update ${\widehat{\bf\alpha}}_j^{(t+1)}$ as a solution to 
\begin{align}\label{ll_alpha_liu} \nonumber
        Q_1^L({\bf \alpha}_j,{\bpsi}^{(t)})=&
        \sum_{i=1}^{n} \I(\Tilde{{Z}}^{(t+1)}_{ij}=1) 
        \log\left\{\pi_j({\bf \omega}_i,{\bf \alpha}) \right\} \\
        & - \frac{1}{2} 
        \left[ \left( -\frac{d_j}{(\lambda_j^*)^{1/2}} \right) 
    {\widehat{\ba}_{j,R}} - (\lambda_j^*)^{1/2} \ba_j
    \right]^\top
    \left[ \left( -\frac{d^*_j}{(\lambda_j^*)^{1/2}} \right) 
    {\widehat{\ba}_{j,R}} - (\lambda_j^*)^{1/2} \ba_j
    \right],
\end{align}
One can easily calculate the gradient of \eqref{ll_alpha_liu} by
 \begin{align}\label{grad_liu_alpha} \nonumber
\nabla_{\ba_j} {\bf Q}^{L}_1(\ba_j,\bpsi^{(t)}) &= 
\sum_{i=1}^{n} 
\left(
\I(\Tilde{{Z}}^{(t+1)}_{ij}=1) - \pi_j({\bf \omega}_i,{\bf \alpha}) 
\right) {\bf \omega}_i  - \lambda_j^* {\ba}_j - d_j^*  {\widehat{\ba}_{j,R}} \\
&= {\bf \omega}^\top {\bf U}_j({\bf \alpha}) 
- \lambda_j^* {\bf\alpha}_j - d_j^*  {\widehat{\ba}_{j,R}} ,
\end{align}
where ${\bf U}_j({\bf \alpha})$ is given by \eqref{u_alpha}. 
Accordingly, the Hessian matrix can be computed by
\begin{align}\label{hess_liu_alpha_vec} \nonumber 
\frac{\partial^2}{\partial \ba_j \ba_l}{\bf Q}^{L}_1(\ba_j,\bpsi^{(t)})  
&= - \sum_{i=1}^{n} \pi_j({\bf \omega}_i,{\bf \alpha})  
\left[ \I(\Tilde{{Z}}^{(t+1)}_{ij}=1) - \pi_j({\bf \omega}_i,{\bf \alpha})  \right] {\bf \omega}_i {\bf \omega}_i^\top - \lambda_j^* \I \\
&= - {\bf \omega}^\top {\bf W}_j({\bf\alpha}^{(t)}) {\bf \omega} 
- \lambda_j^* \I.
\end{align}
where ${\bf W}_j({\bf\alpha}^{(t)})$ comes from \eqref{w_alpha}. 
Let ${\bf V}_j = {\bf\omega}^\top {\bf W}_j({{\bf\alpha}}^{(t)}) {\bf\omega} + \lambda^*_j \I$.  
From \eqref{grad_liu_alpha} and \eqref{hess_liu_alpha_vec}, the IRWLS to update ${\widehat\ba}_j^{(t+1)}$ is given by
\begin{align*}
{\widehat\ba}_j^{(t+1)} &= 
{\ba}_j^{(t)} - \frac{\partial^2}{\partial \ba_j \ba_l}{\bf Q}^{L}_1(\ba_j,\bpsi^{(t)}) 
{\nabla_{\ba_j} {\bf Q}^{L}_2(\bb_j,{\bpsi}^{(t)})} 
|_{\ba_j = {\ba}_j^{(t)}}\\
& = {\ba}_j^{(t)} +  {\bf V}_j^{-1} 
\left\{
{\bf \omega}^\top {\bf U}_j({\ba}^{(t)}) - \lambda^*_j {\ba}_j^{(t)} - d_j^*  {\widehat{\ba}_{j,R}}
\right\}  \\
& = {\bf V}_j^{-1} {\bf \omega}^\top {\bf W}_j({{\bf\alpha}}^{(t)})
\left\{{\bf \omega}  {\ba}_j^{(t)} + [{\bf W}_j({{\bf\alpha}}^{(t)})]^{-1} {\bf U}_j({\ba}^{(t)}) \right\} - d_j^*  {\bf V}_j^{-1} {\widehat{\ba}_{j,R}} \\
& = \left( {\bf \omega}^\top {\bf W}_j({{\bf\alpha}}^{(t)})  {\bf \omega} + \lambda^*_j \I\right)^{-1} 
\left({\bf \omega}^\top {\bf W}_j({{\bf\alpha}}^{(t)})
{\bf v}_j({{\bf\alpha}}^{(t)}) - d_j^*  {\widehat{\ba}_{j,R}}
\right).
\end{align*}


\begin{table}[H]
\caption{\small{The median (M), lower (L) and upper (U) bounds of 90\% CIs for ${\sqrt{\text{MSE}}}$ and Accuracy of the methods in estimation and classification of the FMPRE model with two component regressions when $\rho=0.85$ and $n = 200$.}}
\begin{centering}
\label{2m_n200_r0.85}
\begin{tabular}{ccccccccccc}
\hline 
 &  & \multicolumn{2}{c}{} & \multicolumn{3}{c}{${\sqrt{\text{MSE}}}$} &  & \multicolumn{3}{c}{Accuracy}\tabularnewline
\cline{5-7} \cline{9-11} 
$\phi$ & EM & $\ensuremath{\ensuremath{\boldsymbol{\Psi}}}$ &  & M & L & U &  & M  & L & U\tabularnewline
\hline 
0.90 & LS & ${\bb}$ &  & 0.765 & 0.004 & 87.2 &  & 0.62 & 0.18 & 0.99\tabularnewline
\cline{3-7} 
 &  & ${\ba}$ &  & 0.158 & 0.158 & 27.9 &  &  &  & \tabularnewline
\cline{2-11} 
 & Ridge & ${\bb}$ &  & 0.019 & 0.003 & 11.3 &  & 0.92 & 0.18 & 0.99\tabularnewline
\cline{3-7} 
 &  & ${\ba}$ &  & 0.158 & 0.096 & 21.1 &  &  &  & \tabularnewline
\cline{2-11} 
 & LT & ${\bb}$ &  & 0.011 & 0.003 & 3.6 &  & 0.98 & 0.42 & 0.99\tabularnewline
\cline{3-7} 
 &  & ${\ba}$ &  & 1.146 & 0.146 & 5.1 &  &  &  & \tabularnewline
\hline 
0.95 & LS & ${\bb}$ &  & 0.709 & 0.004 & 128.7 &  & 0.60 & 0.21 & 0.99\tabularnewline
\cline{3-7} 
 &  & ${\ba}$ &  & 0.158 & 0.158 & 30.8 &  &  &  & \tabularnewline
\cline{2-11} 
 & Ridge & ${\bb}$ &  & 0.024 & 0.003 & 21.8 &  & 0.92 & 0.21 & 0.99\tabularnewline
\cline{3-7} 
 &  & ${\ba}$ &  & 0.170 & 0.112 & 22.7 &  &  &  & \tabularnewline
\cline{2-11} 
 & LT & ${\bb}$ &  & 0.014 & 0.003 & 9.7 &  & 0.97 & 0.44 & 0.99\tabularnewline
\cline{3-7} 
 &  & ${\ba}$ &  & 1.129 & 0.149 & 5.1 &  &  &  & \tabularnewline
\hline 
\end{tabular}
\par\end{centering}
\end{table}


\begin{table}[H]
\caption{\small{The median (M), lower (L) and upper (U) bounds of 90\% CIs for ${\sqrt{\text{MSE}}}$ and Accuracy of the methods in estimation and classification of the FMPRE model with two component regressions when $\rho=0.90$ and $n = 200$.}}
\begin{centering}
\label{2m_n200_r0.90}
\begin{tabular}{ccccccccccc}
\hline 
 &  & \multicolumn{2}{c}{} & \multicolumn{3}{c}{${\sqrt{\text{MSE}}}$} &  & \multicolumn{3}{c}{Accuracy}\tabularnewline
\cline{5-7} \cline{9-11} 
$\phi$ & EM & $\ensuremath{\ensuremath{\boldsymbol{\Psi}}}$ &  & M & L & U &  & M  & L & U\tabularnewline
\hline 
0.90 & LS & ${\bb}$ &  & 0.851 & 0.004 & 57.7 &  & 0.62 & 0.18 & 0.98\tabularnewline
\cline{3-7} 
 &  & ${\ba}$ &  & 0.158 & 0.158 & 28.0 &  &  &  & \tabularnewline
\cline{2-11} 
 & Ridge & ${\bb}$ &  & 0.019 & 0.003 & 9.6 &  & 0.90 & 0.18 & 0.99\tabularnewline
\cline{3-7} 
 &  & ${\ba}$ &  & 0.199 & 0.109 & 20.5 &  &  &  & \tabularnewline
\cline{2-11} 
 & LT & ${\bb}$ &  & 0.011 & 0.003 & 4.9 &  & 0.96 & 0.46 & 0.99\tabularnewline
\cline{3-7} 
 &  & ${\ba}$ &  & 1.155 & 0.158 & 5.1 &  &  &  & \tabularnewline
\hline 
0.95 & LS & ${\bb}$ &  & 1.294 & 0.005 & 140.9 &  & 0.55 & 0.17 & 0.99\tabularnewline
\cline{3-7} 
 &  & ${\ba}$ &  & 0.158 & 0.158 & 33.1 &  &  &  & \tabularnewline
\cline{2-11} 
 & Ridge & ${\bb}$ &  & 0.032 & 0.004 & 26.3 &  & 0.90 & 0.17 & 0.99\tabularnewline
\cline{3-7} 
 &  & ${\ba}$ &  & 0.158 & 0.117 & 22.2 &  &  &  & \tabularnewline
\cline{2-11} 
 & LT & ${\bb}$ &  & 0.015 & 0.003 & 9.8 &  & 0.97 & 0.39 & 0.99\tabularnewline
\cline{3-7} 
 &  & ${\ba}$ &  & 1.161 & 0.158 & 5.0 &  &  &  & \tabularnewline
\hline 
\end{tabular}
\par\end{centering}
\end{table}

\newpage


\begin{table}[H]
\caption{\small{The median (M), lower (L) and upper (U) bounds of 90\% CIs for ${\sqrt{\text{MSE}}}$ and Accuracy of the methods in estimation and classification of the FMPRE model with three component regressions when $\rho=0.95$ and $n = 300$.}}
\begin{centering}
\label{3m_n300_r0.95}
\begin{tabular}{cccccccccc}
\hline 
 & \multicolumn{2}{c}{} & \multicolumn{3}{c}{${\sqrt{\text{MSE}}}$} &  & \multicolumn{3}{c}{Accuracy}\tabularnewline
\cline{4-6} \cline{8-10} 
EM & $\ensuremath{\ensuremath{\boldsymbol{\Psi}}}$ &  & M & L & U &  & M  & L & U\tabularnewline
\hline 
LS & ${\bb}$ &  & 0.53 & 0.27 & 10.58 &  & 0.24 & 0.08 & 0.65\tabularnewline
\cline{2-6} 
 & ${\ba}$ &  & 1.48 & 0.33 & 14.72 &  &  &  & \tabularnewline
\hline 
Ridge & ${\bb}$ &  & 0.40 & 0.25 & 3.39 &  & 0.23 & 0.06 & 0.65\tabularnewline
\cline{2-6} 
 & ${\ba}$ &  & 0.72 & 0.25 & 5.77 &  &  &  & \tabularnewline
\hline 
LT & ${\bb}$ &  & 0.33 & 0.20 & 1.05 &  & 0.21 & 0.01 & 0.89\tabularnewline
\cline{2-6} 
 & ${\ba}$ &  & 0.27 & 0.11 & 3.52 &  &  &  & \tabularnewline
\hline 
\end{tabular}
\par\end{centering}
\end{table}


\begin{table}[H]
\caption{\small{The median (M), lower (L) and upper (U) bounds of 90\% CIs for ${\sqrt{\text{MSE}}}$ and Accuracy of the methods in estimation and classification of the heart disease population when $n = 50$.}}
\label{tab:real_50}
\begin{centering}
\begin{tabular}{cccccccccc}
\hline 
 & \multicolumn{2}{c}{} & \multicolumn{3}{c}{${\sqrt{\text{MSE}}}$} &  & \multicolumn{3}{c}{Accuracy}\tabularnewline
\cline{4-6} \cline{8-10} 
EM & $\ensuremath{\ensuremath{\boldsymbol{\Psi}}}$ &  & M & L & U &  & M  & L & U\tabularnewline
\hline 
LS & ${\bb}$ &  & 0.132 & 0.058 & 1.267 &  & 0.96 & 0.91 & 0.99\tabularnewline
\cline{2-6} 
 & ${\ba}$ &  & 0.307 & 0.066 & 5.139 &  & . &  & \tabularnewline
\hline 
Ridge & ${\bb}$ &  & 0.108 & 0.048 & 0.451 &  & 0.97 & 0.92 & 0.99\tabularnewline
\cline{2-6} 
 & ${\ba}$ &  & 0.223 & 0.046 & 1.566 &  &  &  & \tabularnewline
\hline 
LT & ${\bb}$ &  & 0.079 & 0.029 & 0.201 &  & 0.97 & 0.92 & 0.99\tabularnewline
\cline{2-6} 
 & ${\ba}$ &  & 0.165 & 0.056 & 0.441 &  &  &  & \tabularnewline
\hline 
\end{tabular}
\par\end{centering}
\end{table}



\end{document}